\documentclass[a4wide]{autart}
\usepackage{amsmath} 
\usepackage{amssymb}  
\usepackage{psfrag}
\usepackage{amsfonts}
\usepackage{enumerate,cite,latexsym,graphicx}
\newtheorem{theorem}{Theorem}
\newtheorem{lemma}{Lemma}
\newtheorem{corollary}{Corollary}
\newtheorem{definition}{Definition}
\newtheorem{observation}{Observation}
\newtheorem{assumption}{Assumption}
\newtheorem{remark}{Remark}
 \usepackage{color}
\begin{document}

\hyphenation{Qua-dra-tic}

\begin{frontmatter}

\title{A  Kalman Decomposition for Possibly Controllable Uncertain
  Linear Systems\thanksref{footnoteinfo}}
\thanks[footnoteinfo]{This work was supported by the Australian Research Council. Preliminary versions of some of the results of this paper appeared in the 46th IEEE Conference on Decision and Control, New Orleans and the 2008 IFAC World Congress, Seoul.}

\author{Ian R. Petersen$^{*}$} 
\address{$^{*}$School of Engineering and  Information Technology, University of New South Wales at the Australian                    Defence Force Academy, Canberra ACT 2600, Australia (e-mail: i.r.petersen@gmail.com)}            

\begin{abstract}
This paper considers the structure of uncertain linear systems
building on  concepts of  robust unobservability and possible controllability
which were introduced in previous papers. The paper presents a new geometric
characterization of  the possibly controllable states. When combined
with previous geometric results on robust unobservability,  the results of this paper lead to a general Kalman type
decomposition for uncertain linear systems which can be applied to the problem of obtaining reduced order uncertain system models. 
\end{abstract}
\end{frontmatter}
\section{Introduction}
 Controllability and observability are fundamental properties of a
linear system; e.g., see \cite{AM06}. 
This paper is concerned with extending these notions to the case of
uncertain linear systems with the aim of gaining greater understanding
of the structure of uncertain linear systems when applied to problems of reduced order modelling and minimal realization.

One reason for considering the
issue of controllability  for uncertain systems might be to determine if a
robust state feedback controller can be constructed for the system; e.g., see
\cite{PUSB}. In this case, one would be interested in the question of
whether the system is ``controllable'' for all possible values of the
uncertainty; e.g., see \cite{BHA83,PETIJC87,PETMTNS90,CPM91,BM92,SP37}. Similarly, one
reason for considering 
observability for uncertain systems might be to determine if a 
robust state estimator can be constructed for the system; e.g., see
\cite{PSB}. In this case, one would be interested in the question of
whether the system is ``observable'' for all possible values of the
uncertainty; e.g., see \cite{PET00}. However, these questions of robust controllability and robust observability are not the questions being addressed in this paper. 

For the case of linear systems, the notions of controllability and observability are 
central to realization theory; e.g., see \cite{AM06}. For example, it is known that if a
linear system contains unobservable  or uncontrollable states, those states can be
removed in order to obtain a reduced dimension realization of the
system's transfer function. From this point of view, 
a natural extension of the notion of controllability to the case of uncertain systems, would be to consider
``possibly controllable'' states which are controllable for some
possible values of the uncertainty.  This idea was developed in the
paper \cite{PET05B} for the case of uncertain linear 
systems with structured uncertainty subject to averaged integral
quadratic constraints (IQCs). Similarly, a natural extension of the notion
of observability to uncertain systems is to consider
robustly unobservable states which are ``unobservable'' for all
possible values of the uncertainty. This idea was developed in the
papers \cite{PET04A,PET05A}.

This paper builds on concepts of  ``robust unobservability and ``possible
controllability'' developed in the papers \cite{PET04A,PET05B}.
The results presented in the paper  aim to  provide insight into the
structure of uncertain systems as 
it relates to questions of realization theory and reduced dimension modelling for uncertain systems; e.g., see
\cite{BD99,BD04,PET00C}. 

We formally define notions of robust
unobservability and possible controllability in terms of certain
constrained optimization problems. The notion of robust unobservability used in this paper involves
extending the standard linear systems definition of the observability Gramian to the case of
uncertain systems; see also \cite{GM99}. Also,  the notion of possible
controllability used in this paper involves 
extending the standard linear systems definition of the controllability Gramian to the case of
uncertain systems; see also \cite{SG00}.
We then apply the S-procedure (e.g., see \cite{PUSB}) to obtain conditions for robust
unobservability and possible controllability in terms of
unconstrained LQ optimal control problems dependent on Lagrange
multiplier parameters as in \cite{PET04A,PET05B}. 
From this, we  develop a 
geometric  characterization for the set of 
robustly unobservable states (as in \cite{PET05A}) and the set of possibly controllable
states. 
These characterizations imply that the
set of robustly 
unobservable states is in fact a linear subspace. 
Similarly, we show that the set of possibly
controllable states is a linear subspace; see also \cite{BHA83,CPM91,BM92}.
These characterizations lead to a  Kalman type decomposition for the
uncertain systems under consideration; see also \cite{KAL63}, \cite{KAL82} and Theorem 4.3 in Chapter 3 of \cite{AM06}. This decomposition is described in the  four possible cases for 
which an uncertain system
 model can have robustly unobservable states or states which are not
 possibly controllable. These are the cases in which a reduced
 dimension uncertain system model can be obtained which retains the
 same set of  input-output behaviours as the original model. As compared to the previous papers \cite{PET04A,PET05A,PET05B}, the results of this paper enable a complete geometrical picture to be obtained which can be applied to problems of reduced dimension modelling of uncertain linear systems. Also, the results of this paper are much more computationally tractable than the results of the papers \cite{PET04A,PET05B}. The main assumption required in this paper as compared to the previous papers \cite{PET04A,PET05B} is the assumption that the uncertainty is unstructured and described by a single averaged uncertainty constraint. 

The remainder of the paper proceeds as follows. In Section \ref{formulation}, the class of uncertain systems under consideration is introduced and definitions of robust unobservability and possible controllability are given. In Section \ref{existing}, the existing geometrical results on robust observability are summarized. In Sections  \ref{prelim:result}, \ref{riccati}, \ref{main}, our main  results on possible controllability are given. In Section \ref{kalman}, the results are combined to obtain complete Kalman decomposition results and in Section \ref{example}, an illustrative example is given. The paper is concluded in Section \ref{conclusion}.

\section{Problem Formulation}
\label{formulation}
We consider the following  linear time invariant uncertain
system:
\begin{eqnarray}
\label{sys}
\dot{x}(t)& = & Ax(t)+ B_1u(t)
+ B_2\xi(t); \nonumber \\
z(t)& = & C_1x(t)+ D_1u(t); \nonumber \\
y(t)& = & C_2x(t)+ 
 D_2\xi(t)
\end{eqnarray}
where $x\in {\bf R}^n$ is the {\em state},
$y\in {\bf R}^l$ is the {\em measured output},
 $z\in {\bf
R}^{h}$  is the {\em uncertainty output}, $u \in  {\bf R}^{m}$ is the {\em control input}, and
$\xi\in {\bf R}^{r}$ is the {\em uncertainty input}. 

For the system (\ref{sys}), we define the transfer function $G(s)$ to
be the transfer function from the input $\xi(t)$ to the output $y(t)$;
i.e., 
\[
G(s) = C_2(sI-A)^{-1}B_2 + D_2.
\]
Also, we define the transfer function $H(s)$ to
be the transfer function from the input $u(t)$ to the output $z(t)$;
i.e., 
\[
H(s) = C_1(sI-A)^{-1}B_1 + D_1.
\]

\noindent {\em System Uncertainty.}
The uncertainty  in the uncertain system (\ref{sys}) is required to satisfy a certain ``Averaged
Integral Quadratic Constraint''. 

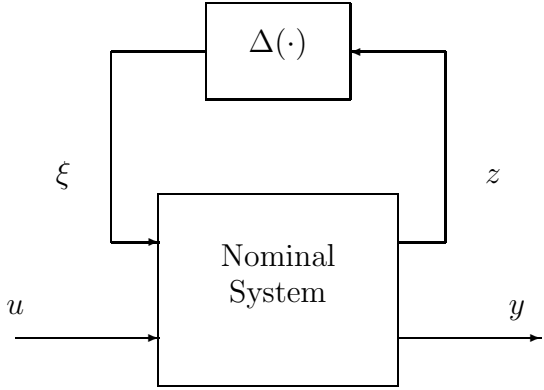
\begin{figure}[htbp]
\begin{center}
\setlength{\unitlength}{3947sp}%
\begingroup\makeatletter\ifx\SetFigFont\undefined%
\gdef\SetFigFont#1#2#3#4#5{%
  \reset@font\fontsize{#1}{#2pt}%
  \fontfamily{#3}\fontseries{#4}\fontshape{#5}%
  \selectfont}%
\fi\endgroup%
\begin{picture}(3451,2424)(1662,-2473)
\thinlines
{\color[rgb]{0,0,0}\put(2701,-2461){\framebox(1500,1200){}}
}%
{\color[rgb]{0,0,0}\put(3001,-661){\framebox(900,600){}}
}%
{\color[rgb]{0,0,0}\put(3001,-361){\line(-1, 0){600}}
\put(2401,-361){\line( 0,-1){1200}}
\put(2401,-1561){\vector( 1, 0){300}}
}%
{\color[rgb]{0,0,0}\put(4201,-1561){\line( 1, 0){300}}
\put(4501,-1561){\line( 0, 1){1200}}
\put(4501,-361){\vector(-1, 0){600}}
}%
{\color[rgb]{0,0,0}\put(1801,-2161){\vector( 1, 0){900}}
}%
{\color[rgb]{0,0,0}\put(4201,-2161){\vector( 1, 0){900}}
}%
\put(3451,-361){\makebox(0,0)[b]{\smash{{\SetFigFont{12}{14.4}{\rmdefault}{\mddefault}{\updefault}{\color[rgb]{0,0,0}$\Delta(\cdot)$}%
}}}}
\put(1801,-2011){\makebox(0,0)[b]{\smash{{\SetFigFont{12}{14.4}{\rmdefault}{\mddefault}{\updefault}{\color[rgb]{0,0,0}$u$}%
}}}}
\put(4951,-2011){\makebox(0,0)[b]{\smash{{\SetFigFont{12}{14.4}{\rmdefault}{\mddefault}{\updefault}{\color[rgb]{0,0,0}$y$}%
}}}}
\put(2101,-1186){\makebox(0,0)[b]{\smash{{\SetFigFont{12}{14.4}{\rmdefault}{\mddefault}{\updefault}{\color[rgb]{0,0,0}$\xi$}%
}}}}
\put(4801,-1186){\makebox(0,0)[b]{\smash{{\SetFigFont{12}{14.4}{\rmdefault}{\mddefault}{\updefault}{\color[rgb]{0,0,0}$z$}%
}}}}
\put(3451,-1711){\makebox(0,0)[b]{\smash{{\SetFigFont{12}{14.4}{\rmdefault}{\mddefault}{\updefault}{\color[rgb]{0,0,0}Nominal}%
}}}}
\put(3451,-1936){\makebox(0,0)[b]{\smash{{\SetFigFont{12}{14.4}{\rmdefault}{\mddefault}{\updefault}{\color[rgb]{0,0,0}System}%
}}}}
\end{picture}%

\end{center}
\caption{Uncertain system block diagram'.}
\label{F0}
\end{figure}

\noindent {\em Averaged Integral Quadratic Constraint.} 
Let the time interval $[0,~T]$, $T > 0$ be given and let $d >0$ be a given positive constant
associated with the system (\ref{sys}); see also \cite{SP5,PET04A,PET05B}.
We will consider sequences of 
uncertainty inputs ${\mathcal S} = \{\xi^1(\cdot), \xi^2(\cdot),
\ldots \xi^q(\cdot)\}$. The number of elements $q$ in any such sequence is
arbitrary. A
sequence of uncertainty functions of the form ${\mathcal S} = \{\xi^1(\cdot),
\xi^2(\cdot), \ldots \xi^q(\cdot)\}$ is an {\em admissible uncertainty
sequence} for the system (\ref{sys}) if the following conditions hold:
Given any 
$\xi^i(\cdot) \in  {\mathcal S}$ and any
corresponding solution $\{x^i(\cdot),\xi^i(\cdot)\}$ to 
(\ref{sys}) defined on $[0,T]$, then 
$\xi^i(\cdot) \in {\bf L}_2[0,T]$, and
\begin{eqnarray}
\label{int}
\frac{1}{q}\sum_{i=1}^q\int_0^T\left(\|\xi^i(t)\|^2
-\|z^i(t)\|^2\right)dt&\leq& d.
\end{eqnarray}
The class of all such admissible uncertainty sequences is denoted ${\bf
\Xi}$. One way in which such uncertainty could be generated is via
unstructured feedback uncertainty as shown in the block diagram  in Fig.
\ref{F0}. 

The averaged IQC uncertainty description was introduced in \cite{SP5} as an approach to uncertainty modelling which gives tight results in the case of structured uncertainty. The paper \cite{PET05B} gives a more detailed explanation concerning the use of the averaged IQC uncertainty description. This paper continues to use the averaged IQC uncertainty description even though it does not consider structured uncertainties since it builds on the results of \cite{PET04A,PET05B} which were derived using the averaged IQC uncertainty description. It should be possible to re-derive the results of \cite{PET04A,PET05B} using the standard rather than averaged IQC uncertainty description such as considered in \cite{SP15}. These results could then be used to obtain results corresponding to the results of this paper in the case of a standard IQC uncertainty description rather than an averaged IQC uncertainty description. 

\begin{definition}
\label{D1}
The {\em robust unobservability function} for the uncertain system
(\ref{sys}), (\ref{int}) defined on the time interval $[0,~T]$ is defined as 
\begin{equation}
\label{LO1}
L_o(x_0,T) \stackrel{\Delta}{=} \sup_{{\mathcal S}\in \Xi} \frac{1}{q}
\sum_{i=1}^q \int_0^T \|y(t)\|^2 dt
\end{equation}
where $x(0) = x_0$ in (\ref{sys}).
\end{definition}

This definition extends the standard definition of the observability
Gramian for linear systems. 

{\em Notation.}
\[
{\mathcal D}\stackrel{\Delta}{=} \{d : d > 0 \}.
\]

\begin{definition}
\label{D2}
A non-zero state $x_0 \in {\bf R}^n$ is said to be {\em robustly
  unobservable} for the 
uncertain system (\ref{sys}), (\ref{int}) defined on the time interval $[0,~T]$ if 
\[
\inf_{d \in {\mathcal D}}L_o(x_0,T) = 0.
\]
The set of all robustly unobservable states for the uncertain
system (\ref{sys}), (\ref{int}) defined on the time interval $[0,~T]$
is referred to as the {\em robustly
  unobservable set} $\mathcal U$; i.e., 
\[
{\mathcal U} \stackrel{\Delta}{=} \left\{x \in {\bf R}^n:
   \inf_{d \in {\mathcal D}} L_o(x,T) = 0 \right\}.
\]
\end{definition}

\begin{definition}
\label{D3}
The {\em possible controllability function} for the uncertain system
(\ref{sys}), (\ref{int}) defined on the time interval $[0,~T]$ is defined as 
\begin{eqnarray}
\label{controllability_function}
&& L_c(x_0,T) \stackrel{\Delta}{=} \nonumber \\
&&\sup_{\epsilon > 0} \inf_{{\mathcal S}\in \Xi} \inf_{{\mathcal U} \in {\bf L}_2^q[0,T]}  \frac{1}{q}
\sum_{i=1}^q \left[\begin{array}{c}\frac{\|x^i(T)\|^2}{\epsilon}\\
+\int_{0}^T
  \|u^i(t)\|^2 dt\end{array}\right] 
\end{eqnarray}
where $x(0) = x_0$ in (\ref{sys}).
\end{definition}

This definition extends the standard definition of the controllability
Gramian for linear systems. In particular, in the special case of systems without uncertainty, this quantity will be infinite for uncontrollable states $x_0$.

\begin{definition}
\label{D4}
A non-zero state $x_0 \in {\bf R}^n$ is said to be {\em possibly
  controllable on $[0,~T]$} for the 
uncertain system (\ref{sys}), (\ref{int}) if 
\[
 \sup_{d \in \mathcal{D}}L_c(x_0,T) < \infty.
\]
\end{definition}

This definition reduces to the definition of controllable states for the special case of systems without uncertainty; e.g., see \cite{AM06}. 

\begin{definition}
\label{D5}
A non-zero state $x_0 \in {\bf R}^n$ is said to be {\em (differentially) possibly
  controllable} for the 
uncertain system (\ref{sys}), (\ref{int}) if it is possibly
  controllable on $[0,~T]$ for all $T > 0$ sufficiently small. 

The set of all differentially possibly controllable states for the uncertain
system (\ref{sys}), (\ref{int}) is referred to as the {\em possibly
  controllable set} $\mathcal{C}$.
\end{definition}

\begin{remark}
It is emphasized in \cite{PET05B} that the notion of possibly controllability for uncertain systems is an extension of the standard notion of controllability in its application to problems of minimal realization. In particular, in the sequel it will be shown that the existence of states which are not possibly controllable in an uncertain system model means that a reduced dimension uncertain system model can be obtained with the same input-output behaviour as the original model. In this sense, states which are not   possibly controllable controllable correspond to uncontrollable states in standard linear systems theory; e.g., see \cite{AM06}. 
\end{remark}

\section{Existing Results on Robust Unobservability}
\label{existing}
In this section, we recall some existing results from \cite{PET05A} giving a geometrical
characterization of robust unobservability. 

For the uncertain system  (\ref{sys}), (\ref{int}) defined on the time interval $[0,~T]$, we define a
function $V_\tau(x_0,T)$ as follows:
\begin{eqnarray}
\label{Vtaudef}
\lefteqn{
V_\tau(x_0,T) \stackrel{\Delta}{=}
}\nonumber \\
&& \inf_{\xi(\cdot)\in {\bf L}_2[0,T]} \int_0^T
\left(\begin{array}{l}
-\|y\|^2  
+\tau \|\xi\|^2
-   \tau \|z\|^2\end{array}\right)dt. \nonumber \\
\end{eqnarray}
Here $\tau \geq 0$ is a  given
constant. 

\[
\bar \Gamma(x_0,T) \stackrel{\Delta}{=} \left\{\begin{array}{l}
\tau:
\tau \geq 0\mbox{ and } V_\tau(x_0,T) >
-\infty \end{array}\right\}.
\]

\begin{assumption}
\label{A1}
For all $x_0 \in
{\bf R}^n$, there exists a constant $\tau \geq 0$ such that $V_\tau(x_0,T) >
-\infty$.
\end{assumption}

{\em Remark:}
The above assumption is a technical assumption required to establish
the  results of \cite{PET05A}. It represents an assumption on the
size of the uncertainty in the system relative to the time interval $[0,~T]$
under consideration. In general, this assumption can always be
satisfied by choosing a sufficiently small $T > 0$. 

\begin{theorem}
\label{T1}
(See \cite{PET05A} for proof).
Consider the uncertain system (\ref{sys}), (\ref{int}) and suppose that
Assumption \ref{A1} is satisfied. Also, suppose that $G(s) \equiv
0$. Then a state $x_0$ is robustly unobservable if and only if it is
an unobservable state for the pair 
$(C_2,A)$. 
\end{theorem} 

\begin{figure}[htbp]
\begin{center}
\includegraphics[width=7cm]{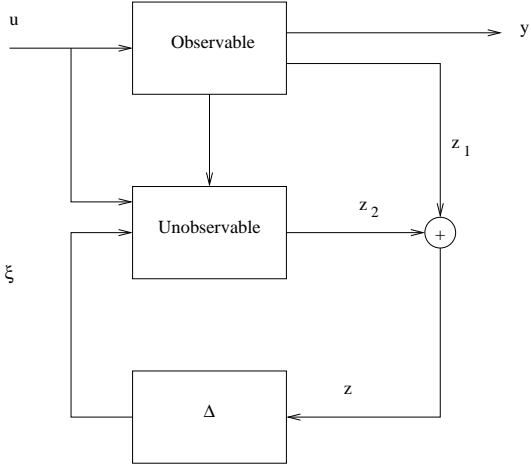}
\end{center}
\caption{Observable-Unobservable decomposition for the uncertain
  system when $G(s) \equiv 0$.}
\label{F1}
\end{figure}

{\em Remark:}
From the above theorem and the fact that $G(s) \equiv 0$, it follows
that we can apply the standard  
Kalman decomposition
to represent the uncertain system as shown in Fig. \ref{F1}. 

Note that in this case,  all of the uncertainty is in
the unobservable subsystem and the coupling between the two
subsystems. 

\begin{theorem}
\label{T2}
(See \cite{PET05A} for proof).
Consider the uncertain system (\ref{sys}), (\ref{int}) and suppose that
Assumption \ref{A1} is satisfied. Also, suppose that $G(s) \not \equiv
0$. Then a state $x_0$ is robustly unobservable if and only if it is
an unobservable state for the pair 
$(\left[\begin{array}{l}C_1\\C_2\end{array}\right],A)$. 
\end{theorem}

\begin{figure}[htbp]
\begin{center}
\includegraphics[width=7cm]{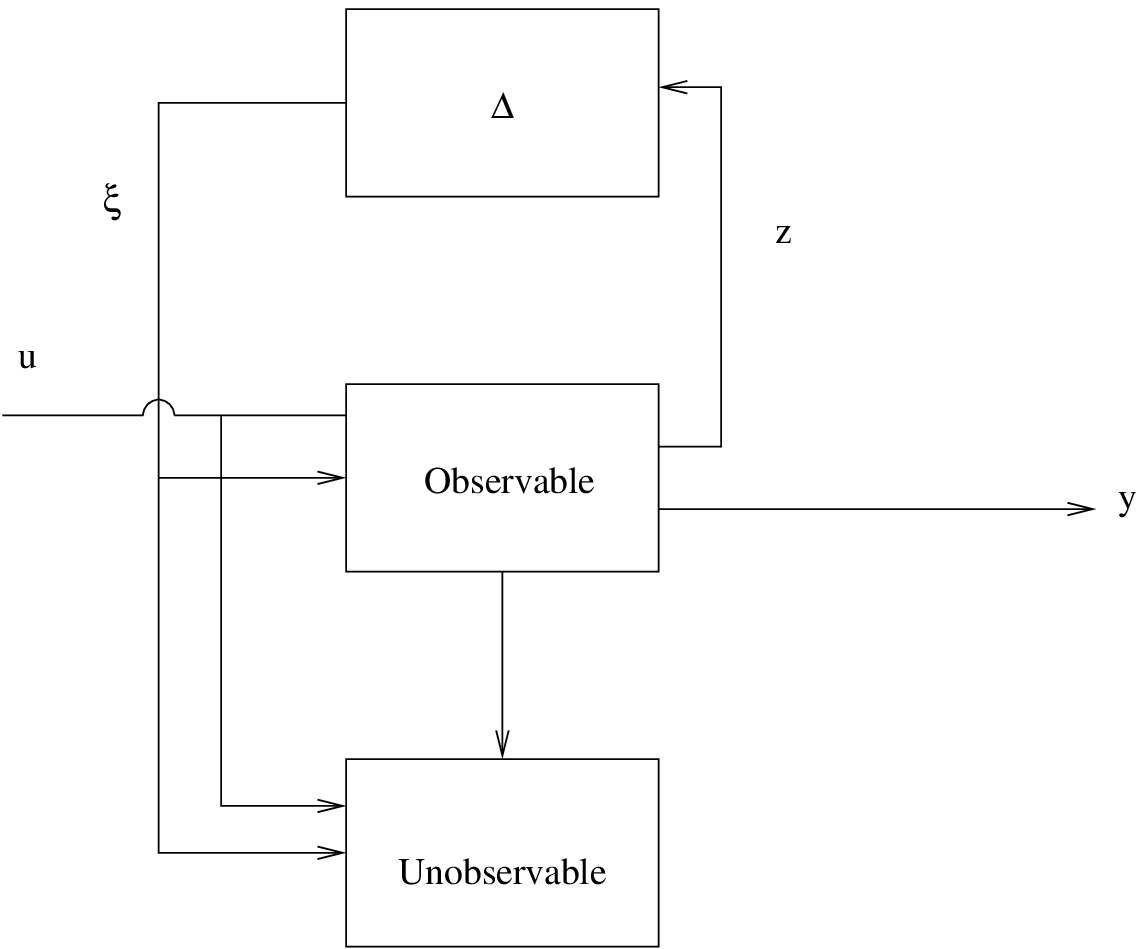}
\end{center}
\caption{Observable-Unobservable decomposition for the uncertain
  system when $G(s) \not \equiv 0$.}
\label{F2}
\end{figure}

{\em Remark:}
The above theorem implies that when $G(s) \not \equiv 0$, the robustly
unobservable set is a linear 
space equal to the unobservable subspace of the pair
$(\left[\begin{array}{l}C_1\\C_2\end{array}\right],A)$. 
From this theorem, it follows that we can apply the standard Kalman
decomposition
to represent the uncertain system as shown in Fig. \ref{F2}.

In this case, all of the uncertainty is in
the observable subsystem or in the coupling between the two
subsystems.

\section{ Preliminary Results on Possible Controllability}
\label{prelim:result}
In this section, we will recall the main results of
\cite{PET05B}  
specialized to the class of uncertain systems with unstructured uncertainty considered in this
paper. 

\subsection{A Family of Unconstrained Optimization Problems.}
For the uncertain system  (\ref{sys}), (\ref{int}) defined on the time interval $[0,~T]$, we define
functions $W^\epsilon_\tau(x_0,\lambda,T)$, $W^\epsilon_\tau(x_0,T)$ and
$W_\tau(x_0,T)$ as follows: 
\begin{eqnarray}
\label{Wtaudef}
W^\epsilon_\tau(x_0,\lambda,T) &\stackrel{\Delta}{=}&
 \inf_{[\xi(\cdot), u(\cdot)] \in {\bf L}_2[\lambda,T]}
 \frac{\|x(T)\|^2}{\epsilon}\nonumber \\ 
&&+\int_{\lambda}^T
\left(\begin{array}{l}
\|u\|^2 
+ \tau \|\xi\|^2
-  \tau \|z\|^2\end{array}\right)dt \nonumber \\
\end{eqnarray}
subject to $x(\lambda) = x_0$;
\begin{eqnarray*}
W^\epsilon_\tau(x_0,T) &\stackrel{\Delta}{=}& W^\epsilon_\tau(x_0,0,T); \\
W_\tau(x_0,T) &\stackrel{\Delta}{=}& \sup_{\epsilon > 0} W^\epsilon_\tau(x_0,T).
\end{eqnarray*} 
Here $\tau \geq 0$ is a  given
constant. 


\subsection{A Formula for the Possible Controllability Function.}

\begin{theorem} (See \cite{PET05B} for proof).
\label{T3}
Consider the uncertain system (\ref{sys}), (\ref{int}) defined on the time interval $[0,~T]$ and
corresponding possible controllability function
(\ref{controllability_function}). Then for any  $x_0 \in {\bf R}^n$,
\begin{eqnarray}
\label{Lc}
L_c(x_0,T) &=& \sup_{\epsilon > 0} \sup_{\tau \geq 0}
\left\{W^\epsilon_\tau(x_0,T) - \tau d\right\}; \nonumber \\
&=& \sup_{\tau \geq 0}\left\{W_\tau(x_0,T) -  
  \tau d\right\}.
\end{eqnarray}
\end{theorem}

\begin{corollary}(See \cite{PET05B} for proof).
\label{C0}
If we define 
\[
\tilde L_c(x_0,T) \stackrel{\Delta}{=} \sup_{d \in \mathcal{D}}L_c(x_0,T)
\]
then
\[
\tilde L_c(x_0,T) = \sup_{\epsilon > 0} \sup_{\tau \geq
  0}W^\epsilon_\tau(x_0,T) = \sup_{\tau  \geq 0}W_\tau(x_0,T).
\]
\end{corollary}

\begin{observation}
\label{O3}
From the above corollary, it follows immediately that 
a non-zero state $x_0 \in {\bf R}^n$ is  (differentially) possibly
  controllable for the 
uncertain system (\ref{sys}), (\ref{int}) if and only if 
\begin{equation}
\label{Wcondition}
 \sup_{\epsilon > 0} \sup_{\tau \geq
  0}W^\epsilon_\tau(x_0,T) =\sup_{\tau \geq 0}W_\tau(x_0,T)< \infty
\end{equation}
for all $T>0$ sufficiently small. 
\end{observation}

\section{Riccati Equation Solution to the Unconstrained Optimization
   Problems}
\label{riccati}
In order to calculate $W^\epsilon_\tau(x_0,\lambda,T)$, 
we note that if $\tau >0$,  and the optimization problem 
(\ref{Wtaudef}) has a finite solution for all initial conditions, then it
can be solved in terms of the following Riccati
differential equation (RDE):
\begin{eqnarray}
\label{rde}
\lefteqn{-\dot{P}^\epsilon =}\nonumber \\
&& A'P^\epsilon+P^\epsilon A\nonumber \nonumber \\
&&-(P^\epsilon B_1-\tau C_1'D_1)\left(I-\tau D_1'D_1\right)^{-1}
(P^\epsilon B_1-\tau C_1'D_1)' \nonumber \\
&&-\frac{P^\epsilon B_2 B_2'P^\epsilon}{\tau} -\tau C_1 C_1';
\hspace{.5cm}P^\epsilon(T) = I/\epsilon 
\end{eqnarray}
which is solved backwards in time. 
\begin{lemma}
\label{L2}
Let $\tau >0$ be such that
\begin{equation}
\label{nonsing1}
I - \tau D_1'D_1 > 0.
\end{equation}
Consider the system
(\ref{sys}) defined on $[0,T]$ and cost functional (\ref{Wtaudef}) with $\lambda \in
[0,T)$. Then  
\[
W^\epsilon_\tau(x_0,\lambda,T) > -\infty ~ \forall x_0 \in {\bf R}^n
\]
if and only if the RDE (\ref{rde}) has a
solution $P^\epsilon_\tau(t)$ defined on $[\lambda,T]$. In this case,
\begin{equation}
\label{Wtau}
W^\epsilon_\tau(x_0,\lambda,T) = x_0'P^\epsilon_\tau(\lambda)x_0.
\end{equation}
\end{lemma}
 {\em Proof.}
This lemma follows directly from a standard LQR optimal control result; e.g., see page 55 of \cite{CA78}. 
$\Box$

In order to calculate $W_\tau(x_0,T)$ using the Riccati equation approach of \cite{PET05B},
we will consider the following RDEs:
\begin{eqnarray}
\label{rdeSeps}
\lefteqn{\dot{S}^\epsilon =}\nonumber \\
&& AS^\epsilon+S^\epsilon A' \nonumber \\
&&-(B_1-\tau S^\epsilon C_1'D_1)\left(I-\tau D_1'D_1\right)^{-1}(B_1-\tau S^\epsilon C_1'D_1)'
 \nonumber \\
&&-\frac{B_2B_2'}{\tau} -\tau S^\epsilon C_1C_1' S^\epsilon;
\hspace{.5cm}S^\epsilon(T) = \epsilon I; 
\end{eqnarray}
\begin{eqnarray}
\label{rdeS}
\lefteqn{\dot{S} =}\nonumber \\
&& AS+S A' \nonumber \\
&&-(B_1-\tau S C_1'D_1)\left(I-\tau D_1'D_1\right)^{-1}(B_1-\tau S C_1'D_1)'
 \nonumber \\
&&-\frac{B_2B_2'}{\tau} -\tau S C_1C_1' S;
\hspace{.5cm}S(T) = 0 
\end{eqnarray}
which are solved backwards in time. 

\begin{theorem} (see \cite{PET05B} for proof.)
\label{T4}
Let $\tau >0$ be such that
$I - \tau D_1'D_1 > 0$. Also suppose there exists an $\epsilon_0
> 0$ such that for all $\epsilon 
\in (0,\epsilon_0)$, all non-zero $x_0 \in {\bf R}^n$ and all $\lambda \in
[0,T]$, then $W^\epsilon_\tau(x_0,\lambda,T) > 0$. Then  for any  $\epsilon 
\in (0,\epsilon_0)$, the Riccati equations
(\ref{rdeSeps}) and (\ref{rdeS}) have solutions $S^\epsilon_\tau(t) > 0$ and
$S_\tau(t) \geq 0$ defined
on $[0,T]$ 
and for any $x_0 \neq 0$
\[
W^\epsilon_\tau(x_0,T) = x_0'\left[S^\epsilon_\tau(0)\right]^{-1}x_0 > 0.
\]
Also, if $S_\tau(0) > 0$ then
\[
W_\tau(x_0,T) = x_0'\left[S_\tau(0)\right]^{-1}x_0 > 0.
\]
Furthermore, if the matrix $S_\tau(0) \geq 0$ is singular and $x_0$ is not
contained within the range space of $S_\tau(0)$, then 
$$W_\tau(x_0,T) =
\infty.$$
\end{theorem}

The following lemma shows that the time interval $[0,T]$ can always be
chosen short enough to guarantee that solutions to the RDEs exist. 

\begin{lemma}
\label{L3}
Let $\epsilon^* > 0$ and $\tau^* > 0$ be given. Then there exists a
sufficiently small
$\tilde T > 0$ such that the RDEs
(\ref{rdeSeps}) and (\ref{rdeS}) both have solutions on $[0,\tilde T]$
and $S^{\epsilon^*}_{\tau*}(t) > 0$. 
\end{lemma}

 {\em Proof.}
This result follows from standard results on differential equations
and the fact that $S^{\epsilon^*}(T) = \epsilon^* I > 0$. 
$\Box$

\begin{lemma}
\label{L4}
Corresponding to the system (\ref{sys}), we consider the dual system:
\begin{eqnarray}
\label{sysdual}
\dot x(t) &=& -A'x(t) + C_1'\xi(t);\nonumber \\
y(t) &=& B_1'x(t) -D_1'\xi(t); \nonumber \\
z(t) &=& B_2'x(t)
\end{eqnarray}
defined on the time interval $[0,T]$, with initial condition
$x(\lambda) = \tilde x_0$ where $\lambda \in [0,T)$. Also, suppose $\epsilon > 0$ and
$\tau > 0$ are such that  the RDEs
(\ref{rdeSeps}) and (\ref{rdeS}) both have solutions on $[0,
T]$. Then, we can write
\begin{eqnarray}
\label{Stauepseqn}
\lefteqn{\tilde x_0'S_\tau^\epsilon(\lambda)\tilde x_0 =}\nonumber \\
&& \sup_{\xi(\cdot) \in {\bf
    L}_2[\lambda,T]} \left\{\begin{array}{l} \epsilon \|x(T)\|^2
+
\\\int_{\lambda}^T\left(\|y\|^2+\frac{1}{\tau}\|z\|^2-\frac{1}{\tau}\|\xi\|^2
\right)dt
\end{array}\right\}\nonumber \\
\end{eqnarray}
and
\begin{eqnarray}
\label{Staueqn}
\lefteqn{\tilde x_0'S_\tau(\lambda)\tilde x_0 =}\nonumber \\
&& \sup_{\xi(\cdot) \in {\bf
    L}_2[\lambda,T]}  
 \int_{\lambda}^T\left(\|y\|^2+\frac{1}{\tau}\|z\|^2-\frac{1}{\tau}\|\xi\|^2
\right)dt
.\nonumber \\
\end{eqnarray}
Furthermore, for any $\lambda \in [0,T)$, we have 
\begin{equation}
\label{Smonotone}
S_\tau^\epsilon(\lambda) \geq S_\tau(\lambda) \geq 0. 
\end{equation}
\end{lemma}

{\em Proof.}
It follows via some straightforward algebraic manipulations that
the RDE (\ref{rdeSeps}) can be re-written as 
\begin{eqnarray}
\label{rdeSeps1}
\lefteqn{\dot{S}^\epsilon =}\nonumber \\
&& AS^\epsilon+S^\epsilon A' \nonumber \\
&&-(S^\epsilon C_1'- B_1D_1')\left(\frac{I}{\tau}- D_1D_1'\right)^{-1}(S^\epsilon C_1'- B_1D_1')'
 \nonumber \\
&&-\frac{B_2B_2'}{\tau} -B_1B_1';
\hspace{.5cm}S^\epsilon(T) = \epsilon I.
\end{eqnarray}
Similarly, the RDE (\ref{rdeS}) can be re-written as
\begin{eqnarray}
\label{rdeS1}
\lefteqn{\dot{S} =}\nonumber \\
&& AS+S A' \nonumber \\
&&-(S C_1'- B_1D_1')\left(\frac{I}{\tau}- D_1D_1'\right)^{-1}(S C_1'- B_1D_1')'
 \nonumber \\
&&-\frac{B_2B_2'}{\tau} -B_1B_1';
\hspace{.5cm}S(T) = 0.
\end{eqnarray}
Then, the formulas (\ref{Stauepseqn}), (\ref{Staueqn}) follow directly
from a standard result on the linear quadratic 
regulator problem; e.g., see page 55 of \cite{CA78}. Also, the first
inequality in (\ref{Smonotone}) follows by comparing
(\ref{Stauepseqn}) and (\ref{Staueqn}), and the second inequality in
(\ref{Smonotone} follows by setting $\xi(\cdot) \equiv 0$ in
(\ref{Staueqn}). 
$\Box$

The following simple linear algebra result will also be useful in the
proof of our main results.

\begin{lemma}
\label{L5}
Let $N$ be a given matrix and let $M>0$ and $\tilde M>0$ be given positive
definite matrices such that
\[
\tilde M = NN'+ M
\]
If the vector $x_0$ can be written as $x_0 = Ny_0$, we have
\[
x_0\tilde M^{-1} x_0 \leq y_0'y_0.
\]
\end{lemma}
{\em Proof.}
It follows from the Matrix Inversion Lemma 
that we can write
\[
I-N'\left(M+NN'\right)^{-1}N = \left(I+N'M^{-1}N\right)^{-1}.
\]
Hence, 
\begin{eqnarray*}
N'\left(M+NN'\right)^{-1}N &=& I-\left(I+N'M^{-1}N\right)^{-1}
\nonumber \\
&\leq & I.
\end{eqnarray*}
Therefore, 
\[
y_0'N'\left(M+NN'\right)^{-1}Ny_0 = x_0'\tilde M^{-1}x_0 \leq y_0'y_0.
\]
This completes the proof of the lemma. 
$\Box$

\section{Main Results on Possible Controllability}
\label{main}
In this section, we present results which provide
a geometric characterization of the differentially possibly controllable states of
the uncertain system (\ref{sys}), (\ref{int}). We first consider the
case in which $H(s) \equiv 0$. 

\begin{theorem}
\label{T5}
Consider the uncertain system (\ref{sys}), (\ref{int}). Also, suppose that $H(s) \equiv
0$. Then a state $x_0$ is differentially possibly controllable if and only if it is
a controllable state for the pair 
$(A,B_1)$. 
\end{theorem}

{\em Proof.}
We first suppose $x_0$ is a differentially possibly controllable state for the
uncertain system (\ref{sys}), (\ref{int}). Hence, using Observation
\ref{O3} it follows that 
\begin{equation}
\label{Wtauineq}
\sup_{\epsilon >  0}\sup_{\tau \geq 0}W_\tau^\epsilon(x_0,T)< \infty
\end{equation}
for all $T>0$ sufficiently small. Now let $\epsilon^* > 0$ and $\tau^* > 0$ be given and
choose $\tilde T >0$ sufficiently small  as in Lemma \ref{L3}. 
Now since $H(s) \equiv 0$, we must have $D_1 = 0$ and it follows from
Lemma \ref{L4} that we can write 
\begin{eqnarray*}
\tilde x_0'S_{\tau^*}(\lambda)\tilde x_0 &=&
 \sup_{\xi(\cdot) \in {\bf
    L}_2[\lambda,\tilde T]}  
 \int_{\lambda}^{\tilde T}\left(\begin{array}{l}
\|y\|^2+\frac{1}{\tau^*}\|z\|^2\\
-\frac{1}{\tau^*}\|\xi\|^2 \end{array}
\right)dt \nonumber \\
&=&  \int_{\lambda}^{\tilde T}\left\|B_1'e^{-A't}\tilde
  x_0\right\|^2dt \nonumber \\
&&+\frac{1}{\tau^*}\sup_{\xi(\cdot) \in {\bf
    L}_2[\lambda,\tilde T]}  
 \int_{\lambda}^{\tilde T}\left(\begin{array}{l}
\|z\|^2
-\|\xi\|^2 \end{array}
\right)dt \nonumber \\
&=& \tilde x_0'W_c(\lambda,\tilde T)\tilde x_0 
+ \frac{1}{\tau^*} \tilde x_0'Q(\lambda,\tilde T)\tilde x_0
\end{eqnarray*}
where
\begin{equation}
\label{Qdef}
\tilde x_0'Q(\lambda,\tilde T)\tilde x_0 = 
\sup_{\xi(\cdot) \in {\bf
    L}_2[\lambda,\tilde T]}  
 \int_{\lambda}^{\tilde T}\left(\begin{array}{l}
\|z\|^2
-\|\xi\|^2 \end{array}
\right)dt \geq 0 
\end{equation}
and 
\[
W_c(\lambda,\tilde T) = \int_{\lambda}^{\tilde T}e^{-At}B_1B_1'e^{-A't}dt
\]
is the controllability Gramian for the pair $(A,B_1)$; e.g., see
\cite{AM06}. 
From this, we can conclude that 
\begin{equation}
\label{Stauexpression}
S_{\tau}(\lambda) = W_c(\lambda,\tilde T) +  \frac{1}{\tau}
Q(\lambda,\tilde T) 
\end{equation}
is monotone decreasing as $\tau$ increases and hence, the RDE
(\ref{rdeS}) does not have a finite escape  
time on $[0,\tilde T]$ for all $\tau \geq \tau^*$. Furthermore, it
follows from the continuity of solutions to the RDEs  (\ref{rdeS}) and  (\ref{rdeSeps}) that for all $\tau
\geq \tau^*$, there exists a $\epsilon \in (0,\epsilon^*)$ sufficiently
small such that the RDE (\ref{rdeSeps}) has a solution
$S_{\tau}^\epsilon(t)$ on $[0,\tilde T]$. 
We now observe that $S_{\tau}^\epsilon(\lambda) > S_{\tau}(\lambda)
\geq 0$ for all $\lambda \in [0,\tilde T]$. Indeed, given any non-zero
$\tilde
x_0 \in {\bf R}^n$, it follows from  (\ref{Stauepseqn}) and
(\ref{Staueqn}) that
\begin{equation}
\label{Stauepsineq}
\tilde x_0'S_{\tau}^\epsilon(\lambda)\tilde x_0\geq \tilde x_0'S_{\tau}(\lambda)\tilde x_0
+ \epsilon \|x^*(T)\|^2
\end{equation}
where $x^*(t)$ is the solution to (\ref{sysdual}) with initial
condition $x(\lambda) = \tilde x_0$ and input $\xi^*(\cdot)$ which
achieves the supremum in (\ref{Staueqn}). Furthermore, since $S_{\tau}(t)$ the
solution to RDE (\ref{rdeS}) exists on
$[0,\tilde T]$, it follows by a standard result on linear quadratic
optimal control (e.g., see \cite{CA78}) that $\xi^*(\cdot)$ is defined
by  the following state feedback control law for (\ref{sysdual})
\[
\xi^*(t) = -\tau C_1 S_{\tau}(t)x^*(t).
\]
Then, we can write $x^*(T) = \Phi(\tilde T,\lambda)\tilde x_0$ where
$\Phi(\tilde T,\lambda)$ is the state transition matrix for the
closed loop system 
\[
\dot x = \left(-A' - \tau C_1'C_1 S_{\tau}(t)\right)x.
\]
Hence, it follows from (\ref{Stauepsineq}) that 
\[
\tilde x_0'S_{\tau}^\epsilon(\lambda)\tilde x_0\geq \tilde x_0'S_{\tau}(\lambda)\tilde x_0
+ \epsilon \|\Phi(\tilde T,\lambda)\tilde x_0\|^2 > \tilde x_0'S_{\tau}(\lambda)\tilde x_0.
\]
Thus, we can conclude that $S_{\tau}^\epsilon(\lambda) > S_{\tau}(\lambda)
\geq 0$ for all $\lambda \in [0,\tilde T]$. Also, it follows from Lemma
\ref{L4}, that for any $\lambda \in [0,\tilde T)$ that
$S_\tau^\epsilon(\lambda)$ is monotone decreasing as $\epsilon
\rightarrow 0$. 

We have now established that given any $\tau \geq \tau^*$, there exists
an $\epsilon \in (0,\epsilon^*)$ such that $S_{\tau}^\epsilon(t)$ the
solution to (\ref{rdeSeps}) exists on $[0,\tilde T]$ and
$S_{\tau}^\epsilon(t) > 0$ for all $t \in [0,\tilde T]$. From this, it
follows that $P_{\tau}^\epsilon(t) =
\left[S_{\tau}^\epsilon(t)\right]^{-1} > 0$ is the solution to
(\ref{rde}) on $[0,\tilde T]$. Therefore, it follows from Theorem
\ref{T4} that given any $x_0 \in   {\bf R}^n$, then we can write
\[ 
W^\epsilon_\tau(x_0,\tilde T) = x_0'\left[S^\epsilon_\tau(0)\right]^{-1}x_0. 
\]

Now we return to the inequality (\ref{Wtauineq}) for our
differentially possibly controllable state $x_0$ and conclude that
there exists a constant $M \geq 0$ such that given any integer $k \geq
k_0 \geq 
\tau^*$, there exists an $\epsilon_k \in  (0,\epsilon^*)$ such that 
\begin{equation}
\label{Wepsbound}
W^\epsilon_k(x_0,\tilde T) =
x_0'\left[S^{\epsilon_k}_k(0)\right]^{-1}x_0 \leq M.
\end{equation}
Also, we can assume without loss of generality that $\epsilon_k
\rightarrow 0$ as $k \rightarrow \infty$. We now define a sequence
$\{y_0^k\}_{k=k_0}^\infty$ as
\[
y_0^k = \left[S^{\epsilon_k}_k(0)\right]^{-\frac{1}{2}}x_0.
\]
Hence, we have 
\begin{equation}
\label{y0k}
x_0 = \left[S^{\epsilon_k}_k(0)\right]^{\frac{1}{2}}y_0^k~~\forall ~k
\geq k_0
\end{equation}
and therefore it follows from (\ref{Wepsbound}) that 
\[
\|y_0^k\|^2 \leq M ~~ ~~\forall ~k
\geq k_0.
\]
From this, we can conclude that the sequence
$\{y_0^k\}_{k=k_0}^\infty$ has a convergence subsequence $\{\tilde
y_0^k\}_{k=k_0}^\infty$:
\[
\tilde y_0^k \rightarrow \bar y_0.
\]
Now, using the fact that for any $\tau \geq \tau*$, then
$S^\epsilon_\tau(0) \rightarrow S_\tau(0)$ as $\epsilon \rightarrow
0$, combined with (\ref{Stauexpression}), it follows from (\ref{y0k})
that we can write
\[
x_0 = \left[W_c(0,\tilde T)\right]^{\frac{1}{2}}\bar y_0.
\]
That is, $x_0$ is in the range space of the controllability Gramian
and hence, $x_0$ is a controllable state for the pair $(A,B_1)$. 

Conversely, suppose $x_0$ is a controllable state for the pair
$(A,B_1)$. Let $\epsilon^* > 0$ and $\tau* > 0$ be any positive
constants. Also let $\tilde T > 0$ be any sufficiently small time
horizon chosen as in Lemma \ref{L3}. Then as above, given any $\tau
\geq \tau^*$, $S_{\tau}(t)$ the
solution to (\ref{rdeS}) exists and is positive semidefinite on $[0,\tilde
T]$ and satisfies (\ref{Stauexpression}). Also, it follows from
(\ref{Stauexpression}) that for all $\tau > 0$, the 
solution to (\ref{rdeS}) exists and is positive semidefinite on $[0,\tilde
T]$.  Furthermore  also as above, given any $\tau
> 0$, there exists a sufficiently small $\epsilon \in
(0,\epsilon^*)$ such that $S_{\tau}^\epsilon(t)$ the
solution to (\ref{rdeSeps}) exists and is positive definite on $[0,\tilde
T]$. Moreover, we have $S_{\tau}^\epsilon(\lambda) > S_{\tau}(\lambda)
\geq 0$  and $S_{\tau}^\epsilon(\lambda) \rightarrow S_{\tau}(\lambda)$
as $\epsilon \rightarrow 0$ for all $\lambda \in [0,\tilde T]$. Hence,
using (\ref{Stauexpression}), we can write
\begin{equation}
\label{Stauexp1}
S_{\tau}^\epsilon(0) = W_c(0,\tilde T)+\Phi^\epsilon
\end{equation}
where $\Phi^\epsilon = S_{\tau}^\epsilon(0)- S_{\tau}(0)+ \frac{1}{\tau}
Q(0,\tilde T) > 0$ and $Q(0,\tilde T) \geq 0$ is defined as in 
(\ref{Qdef}). 

Now using the fact that $x_0$ is a controllable state for the pair
$(A,B_1)$, it follows that we can write 
\[
x_0 = \left[W_c(0,\tilde T)\right]^{\frac{1}{2}} y_0
\]
for some vector $y_0$ where $W_c(0,\tilde T)$ is the controllability
Gramian for the pair $(A,B_1)$. Thus, using (\ref{Stauexp1}) and Lemma
\ref{L5}, we conclude that 
\begin{equation}
\label{Wtauepsbound1}
W_\tau^\epsilon(x_0,\tilde T)=
x_0'\left[S^\epsilon_\tau(0)\right]^{-1}x_0 \leq y_0'y_0. 
\end{equation}
Now for fixed $\tau > 0$, it follows from the definition that
$W_\tau^\epsilon(x_0,\tilde T)$ is monotonically increasing as
$\epsilon \rightarrow 0$. Also, (\ref{Wtauepsbound1}) holds for all
sufficiently small $\epsilon > 0$. Thus, we must have 
\begin{equation}
\label{Wtauepsbound2}
W_\tau(x_0,\tilde T) =\sup_{\epsilon > 0}W_\tau^\epsilon(x_0,\tilde T)\leq y_0'y_0 
\end{equation}
for all $\tau > 0$.

We now consider the case of $\tau = 0$. In this case, 
\begin{eqnarray*}
W^\epsilon_0(x_0,0,\tilde T) &=&
 \inf_{[\xi(\cdot), u(\cdot)] \in {\bf L}_2[0,T]}
 \frac{\|x(T)\|^2}{\epsilon}\nonumber \\ 
&&+\int_{0}^T
\left(
\|u\|^2 
\right)dt. \nonumber \\
\end{eqnarray*}
Now since $x_0$ is a controllable state for the pair
$(A,B_1)$, it follows that there exists a control $u^*(\cdot)$ defined
on $[0,\tilde T]$ such that with $\xi(\cdot) \equiv 0$, then $x(T) =
0$. Hence,
\[
W^\epsilon_0(x_0,0,\tilde T) \leq \int_{0}^T
\left(
\|u^*\|^2 
\right)dt
\]
for all $\epsilon > 0$. Therefore, we have
\[
W_0(x_0,\tilde T) =\sup_{\epsilon > 0}W_0^\epsilon(x_0,\tilde T)\leq
\int_{0}^T
\left(
\|u^*\|^2 
\right)dt.
\]

We have now shown that 
\[
W_\tau(x_0,\tilde T) < \infty
\]
for all $\tau \geq 0$ and for all $\tilde T > 0$ sufficiently
small. Thus, using Observation \ref{O3}, we can conclude that $x_0$ is
a
differentially possibly controllable state. This completes the proof.
$\Box$

\begin{figure}[htbp] 
\begin{center}
\includegraphics[width=7cm]{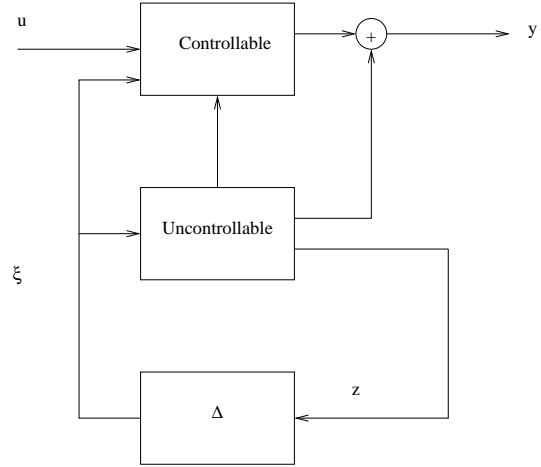}
\end{center}
\caption{Control-Uncontrollable decomposition for the uncertain
  system when $H(s) \equiv 0$.}
\label{F3}
\end{figure}

{\em Remark:}
The above theorem implies that when $H(s) \equiv 0$  the possibly
controllable set is a linear  
space equal to the controllable subspace of the pair $(A,B_1)$. 
From the above theorem and the fact that $H(s) \equiv 0$, it follows
that we can apply the standard  
Kalman decomposition
to represent the uncertain system as shown in Fig. \ref{F3}.

In this case, we only have uncertainty in the uncontrollable
subsystem or in the coupling between the two subsystems.

We now consider the
case in which $H(s) \not \equiv 0$. 

\begin{theorem}
\label{T6}
Consider the uncertain system (\ref{sys}), (\ref{int}) and suppose
that $H(s) \not \equiv
0$. Then, a state $x_0$ is differentially possibly
controllable if and only if  $x_0$ is
a controllable state for the pair 
$(A,[B_1~ B_2])$. 
\end{theorem}

{\em Proof.}
 Suppose $x_0$ is a differentially possibly controllable state for the
uncertain system (\ref{sys}), (\ref{int}). Hence, using Observation
\ref{O3} it follows that 
\begin{equation}
\label{Wtauineq}
\sup_{\epsilon >  0}\sup_{\tau \geq 0}W_\tau^\epsilon(x_0,T)< \infty
\end{equation}
for $T>0$ sufficiently small. Setting $\tau = 0$, it follows that
there exists a constant $M > 0$ such that 
\[
 \inf_{[\xi(\cdot), u(\cdot)] \in {\bf L}_2[0,T]}
 \frac{\|x(T)\|^2}{\epsilon}
+\int_{0}^T
\|u\|^2 dt \leq M~~\forall \epsilon > 0
\]
where the inf is defined for the system (\ref{sys}) with initial
condition $x(0) = x_0$. From this it follows that
\[
 \inf_{[\xi(\cdot), u(\cdot)] \in {\bf L}_2[0,T]}
 \|x(T)\|^2
\leq \epsilon M~~\forall \epsilon > 0
\]
and hence,
\[
 \inf_{[\xi(\cdot), u(\cdot)] \in {\bf L}_2[0,T]}
 \|x(T)\|^2
=0.
\]
Therefore, the state $x_0$ must be a controllable state for the pair
$(A,[B_1~ B_2])$. 

We now suppose  the state $x_0$ is  a controllable state for the pair
$(A,[B_1~ B_2])$ and show that $x_0$ is a differentially possibly controllable state for the
uncertain system (\ref{sys}), (\ref{int}).
In order to prove that the state  $x_0$ is possibly controllable, we
must show that for all $T>0$
sufficiently small $\sup_{\tau \geq 0}W_\tau(x_0,T)< \infty$. In order to
show this, we let $T
>0$ be given and establish the following claim:

{\bf Claim.} For the system (\ref{sys}), there exists an input pair  $\{u^*(\cdot),\xi^*(\cdot)\}$ defined on $[0,T]$ such that $x(0) = x_0$, $x(T) = 0$ and 
\[
\int_{0}^T\left(\begin{array}{l}\|\xi^*\|^2
-  \|z^*\|^2\end{array}\right)
dt \leq 0.
\]

To establish this claim, we first suppose that the standard Kalman
decomposition is applied to the pair $(A,B_1)$ to decompose it into
controllable and uncontrollable subsystems. That is, we can assume
without loss of generality that the system (\ref{sys}) is such that
the matrices $A$, $B_1$, $B_2$, $C_1$ and the
vector $x$ are of
the form 
\begin{eqnarray}
\label{decomposition}
A &=& \left[\begin{array}{ll}A_{11} & A_{12} \\ 0 &
    A_{22}\end{array}\right];
~~B_1 =
\left[\begin{array}{ll}B_{11}\\0\end{array}\right];\nonumber \\
B_2 &=& \left[\begin{array}{l}B_{21}\\B_{22}\end{array}\right];~~
C_1=\left[\begin{array}{ll}C_{11} & C_{12}
    \end{array}\right];\nonumber \\
x&=&\left[\begin{array}{l}x_{1} \\ x_2\end{array}\right]
\end{eqnarray}
where the pair $(A_{11},B_{11})$ is controllable. 

Now consider an input pair $\{\bar u(\cdot),\bar \xi(\cdot)\}$ defined on $[0,\frac{T}{3}]$ such that $x(0) 
= x_0$ and $x(\frac{T}{3}) = 0$.  Such an input pair
exists due to our assumption that  $x_0$ is  a controllable state for the pair
$(A,[B_1~ B_2])$. Then, we can write
\[
J_1 = \int_{0}^{\frac{T}{3}}\left(\begin{array}{l}\|\bar \xi\|^2
-  \|\bar z\|^2\end{array}\right)
dt < \infty.
\]

Now for $t \in (\frac{T}{3},\frac{2T}{3}]$, consider the input pair
$\{\hat u(\cdot),\hat \xi(\cdot)\}$ defined so that $\hat \xi(\cdot)
\equiv 0$ and so that $\hat u(\cdot)$ is such that the corresponding
uncertainty output $\hat z(\cdot) \not \equiv 0$. Such an input $\hat
u(\cdot)$ exists since we have assumed that $H(s) \not \equiv
0$. Then, we let
\[
\gamma = \int_{\frac{T}{3}}^{\frac{2T}{3}}\|\hat z\|^2
dt > 0.
\]
Also, 
since $x(\frac{T}{3}) = 0$ and  $\hat \xi(t)
= 0$ for $t \in (\frac{T}{3},\frac{2T}{3}]$, it follows from
(\ref{decomposition}) that $x_2(t) = 0$ for $t \in
(\frac{T}{3},\frac{2T}{3}]$. 

Now for $t \in (\frac{2T}{3},T]$, consider the input pair
$\{\check u(\cdot),\check \xi(\cdot)\}$ defined so that $\check \xi(\cdot)
\equiv 0$ and so that $\hat u(\cdot)$ is such that $x_1(T) = 0$. Such
an input $\check u(\cdot)$ exists since we have assumed that the pair
$(A_{11},B_{11})$ is controllable. Also, 
since $x_2(\frac{2T}{3}) = 0$ and  $\check \xi(t)
= 0$ for $t \in (\frac{2T}{3},T]$, it follows from
(\ref{decomposition}) that $x_2(t) = 0$ for $t \in
(\frac{2T}{3},T]$. We let $\check z(t)$ denote the corresponding
uncertainty output for $t \in (\frac{2T}{3},T]$. 

We now consider an  input pair $\{u^*(\cdot),\xi^*(\cdot)\}$ defined as
follows:
\begin{eqnarray*}
u^*(t)&=& \left\{\begin{array}{ll}\bar u(t) & \mbox{ for } t \in
    [0,\frac{T}{3}]; \\
\hat u(t)& \mbox{ for } t \in
    (\frac{T}{3},\frac{2T}{3}]; \\
\check u(t)& \mbox{ for } t \in
    (\frac{2T}{3},T];
\end{array}\right.\\
\xi^*(t)&=& \left\{\begin{array}{ll}\bar \xi(t) & \mbox{ for } t \in
    [0,\frac{T}{3}]; \\
0 & \mbox{ for } t \in
    (\frac{T}{3},T].
\end{array}\right.\\
\end{eqnarray*}
It follows from this construction that the pair
$\{u^*(\cdot),\xi^*(\cdot)\}$ gives $x(T) = 0$ and 
\begin{eqnarray*}
\lefteqn{\int_{0}^T\left(\|\xi^*\|^2
-  \|z^*\|^2\right)
dt }\nonumber \\
&=& \int_{0}^{\frac{T}{3}}\left(\|\bar \xi\|^2
-  \|\bar z\|^2\right)dt 
-  \int_{\frac{T}{3}}^{\frac{2T}{3}}\|\hat z\|^2dt \\
&&-  \int_{\frac{2T}{3}}^{T}\|\check z\|^2dt \\
&\leq & J_1 - \gamma. 
\end{eqnarray*}

We now let $\mu > 0$ be a scaling parameter and introduce a modified
input pair $\{u^*(\cdot),\xi^*(\cdot)\}$ defined as
follows:
\begin{eqnarray*}
u^*(t)&=& \left\{\begin{array}{ll}\bar u(t) & \mbox{ for } t \in
    [0,\frac{T}{3}]; \\
\mu \hat u(t)& \mbox{ for } t \in
    (\frac{T}{3},\frac{2T}{3}]; \\
\mu \check u(t)& \mbox{ for } t \in
    (\frac{2T}{3},T];
\end{array}\right.\\
\xi^*(t)&=& \left\{\begin{array}{ll}\bar \xi(t) & \mbox{ for } t \in
    [0,\frac{T}{3}]; \\
0 & \mbox{ for } t \in
    (\frac{T}{3},T].
\end{array}\right.\\
\end{eqnarray*}
It is straightforward to verify that this input pair also leads to
$x(T) = 0$ and  
\[
\int_{0}^T\left(\|\xi^*\|^2
-  \|z^*\|^2\right)
dt \leq  J_1 - \mu^2 \gamma. 
\]
Letting, 
\[
\mu = \sqrt{\frac{J_1}{\gamma}}
\]
it follows that
\[
\int_{0}^T\left(\|\xi^*\|^2
-  \|z^*\|^2\right)
dt \leq 0
\]
and hence, the conditions of the claim are satisfied. This completes
the proof of the claim. 

We now use this claim to complete the proof. Indeed, for any  $\tau
\geq 0$ and $\epsilon > 0$, we have 
\begin{eqnarray}
\label{W0}
W^\epsilon_\tau(x_0,T) &=& \inf_{[\xi(\cdot), u(\cdot)] \in {\bf L}_2[0,T]}
 \frac{\|x(T)\|^2}{\epsilon}
\nonumber \\ &&
+\int_{0}^T
\left(\begin{array}{l}
\|u\|^2 
+ \tau \|\xi\|^2
-  \tau \|z\|^2\end{array}\right)
dt \nonumber \\
&\leq& \int_{0}^T\left(\begin{array}{l}
\|u^*\|^2 
+ \tau \|\xi^*\|^2
-  \tau \|z^*\|^2\end{array}\right)
dt \nonumber \\
\end{eqnarray}
where the input pair $\{u^*(\cdot),\xi^*(\cdot)\}$ is constructed
using the above claim such that $x(0)
= x_0$ and $x(T) = 0$ and
\[
\int_{0}^T\left(\begin{array}{l}\|\xi^*\|^2
-  \|z^*\|^2\end{array}\right)
dt \leq 0.
\]
Also, $z^*(\cdot)$
is the corresponding uncertainty output for the system (\ref{sys}). 
Since $\epsilon > 0$ was arbitrary, it follows from
(\ref{W0}) that 
\begin{eqnarray}
\label{Wbound0}
W_\tau(x_0,T) &=& \sup_{\epsilon > 0}
W^\epsilon_\tau(x_0,T) \nonumber \\
&\leq& \int_{0}^T
\|u^*\|^2 dt
+ \tau  \int_{0}^T\left(\begin{array}{l}\|\xi^*\|^2
-  \|z^*\|^2\end{array}\right)
dt \nonumber \\
&\leq & \int_{0}^T
\|u^*\|^2 dt
\end{eqnarray}
for all $\tau \geq 0$. Thus, we can conclude that 
$$\sup_{\tau \geq 0}W_\tau(x_0,T)< \infty.$$
 Since, $T > 0$ was
arbitrary, it follows from Observation \ref{O3} that $x_0$ is
differentially possibly  
controllable. This completes the proof of the theorem.
\hfill $\Box$ 

\begin{figure}[htbp]
\begin{center}
\includegraphics[width=7cm]{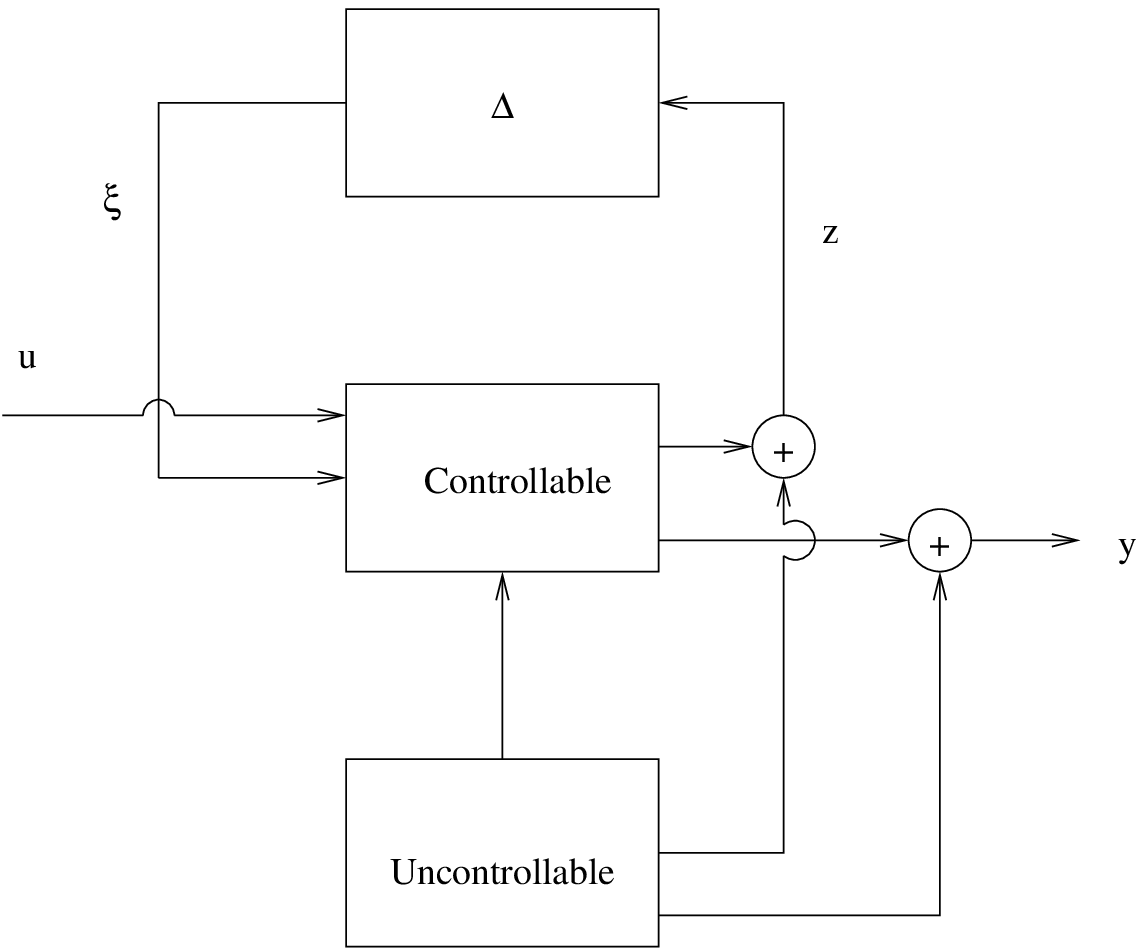}
\end{center}
\caption{ Control-Uncontrollable decomposition for the uncertain
  system when $H(s) \not \equiv 0$.}
\label{F4}
\end{figure}

{\em Remark:}
The above theorem implies that when $H(s) \not \equiv 0$  the possibly 
controllable set is a linear 
space equal to the controllable subspace of the pair $(A,[B_1~B_2])$. 
From the above theorem, it follows
that we can apply the standard  
Kalman decomposition
to represent the uncertain system as shown in Fig. \ref{F4}.

In this case, we  only have uncertainty in the controllable
subsystem or in the coupling between the two subsystems. 

\section{Kalman Decompositions}
\label{kalman}
\begin{figure}[htbp]
\begin{center}
\includegraphics[width=8cm]{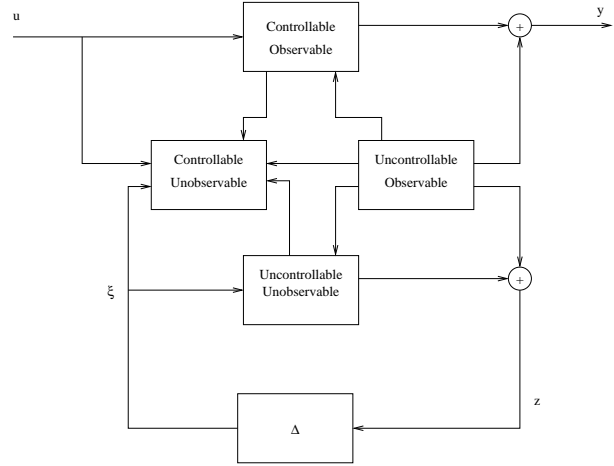}
\end{center}
\caption{Kalman decomposition for the uncertain
  system when $G(s) \equiv 0$, $H(s) \equiv 0$.}
\label{F5}
\end{figure}

We can now combine the results of Theorems \ref{T1}, \ref{T2},
\ref{T5}, and \ref{T6}
to obtain a complete Kalman decomposition for 
the uncertain system in the following cases:

{\bf Case 1}
$G(s) \equiv 0$, $H(s) \equiv 0$. In this case, we  apply the
standard Kalman decomposition to the triple $(C_2,A,B_1)$ to obtain
the situation as illustrated in the  block diagram shown in
Fig. \ref{F5}. 
 
This situation corresponds to uncertainty only in the
uncontrollable-unobservable block. Also there is uncertainty in the coupling between
uncontrollable-observable block and the uncontrollable-unobservable
block. Furthermore, there is uncertainty in the coupling between the uncontrollable-unobservable
block and the controllable-unobservable block. 

\begin{figure}[htbp]
\begin{center}
\includegraphics[width=8cm]{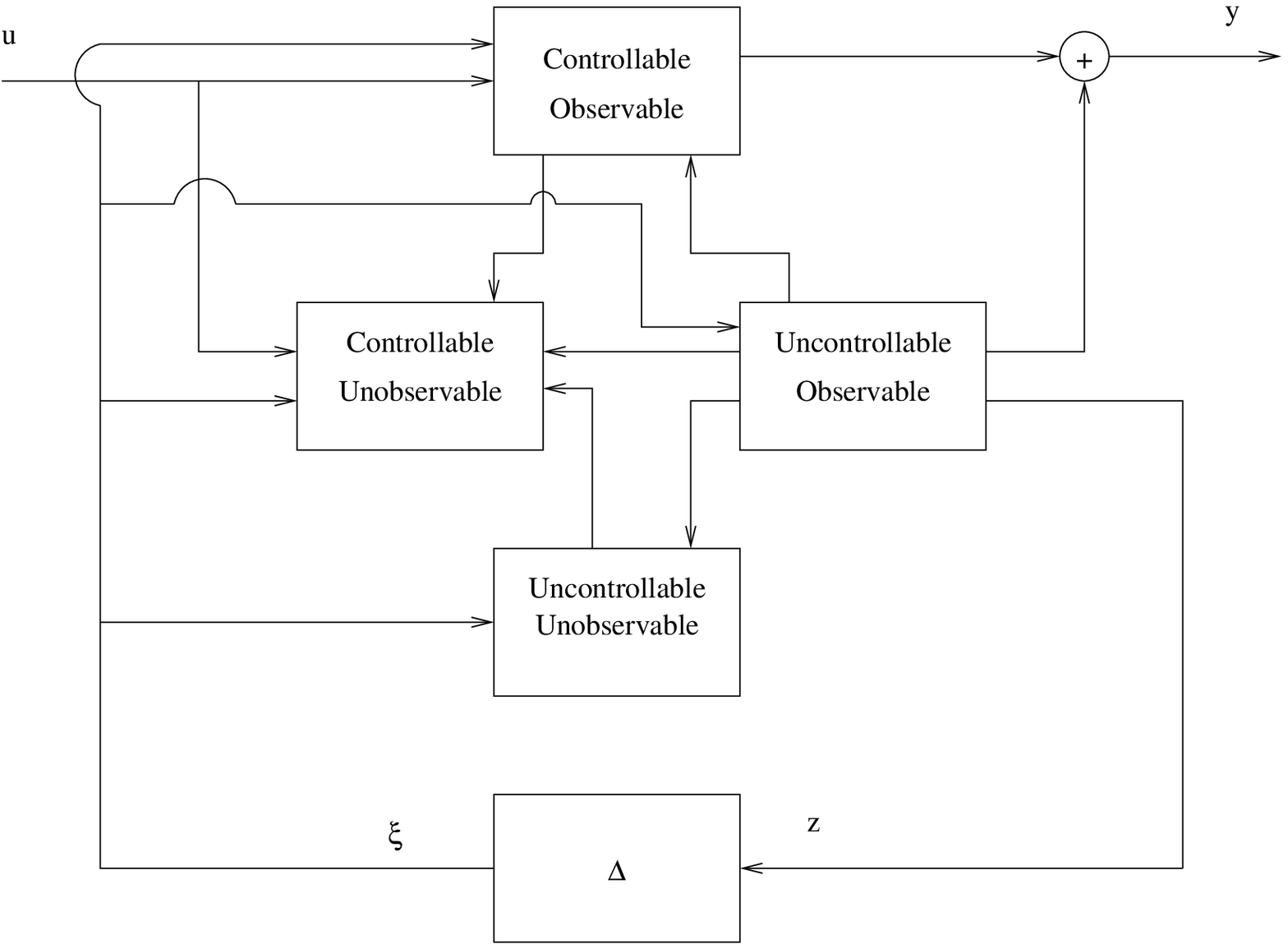}
\end{center}
\caption{Kalman decomposition for the uncertain
  system when $G(s) \not \equiv 0$, $H(s) \equiv 0$.}
\label{F6}
\end{figure}

{\bf Case 2}
$G(s) \not \equiv 0$, $H(s) \equiv 0$. In this case, we  apply the
standard Kalman decomposition to the triple
$$(\left[\begin{array}{l}C_1\\C_2\end{array}\right],A,B_1)$$
 to obtain
the situation as illustrated in the  block diagram shown in Fig.
 \ref{F6}. 

This situation corresponds to uncertainty only in the
uncontrollable-observable block. Also there is uncertainty in the coupling between
uncontrollable-observable block and the uncontrollable-unobservable
block. Furthermore, there is uncertainty in the coupling between the
uncontrollable observable
block and the controllable-unobservable block. As well, there is
uncertainty in the coupling between the uncontrollable-observable 
block and the controllable-observable block.

Note that in order to guarantee that the condition $H(s) \equiv 0$ we
needed to make a further restriction on the controllable observable
block in the above diagram so that  in fact it only has an output
$y$.

\begin{figure}[htbp]
\begin{center}
\includegraphics[width=8cm]{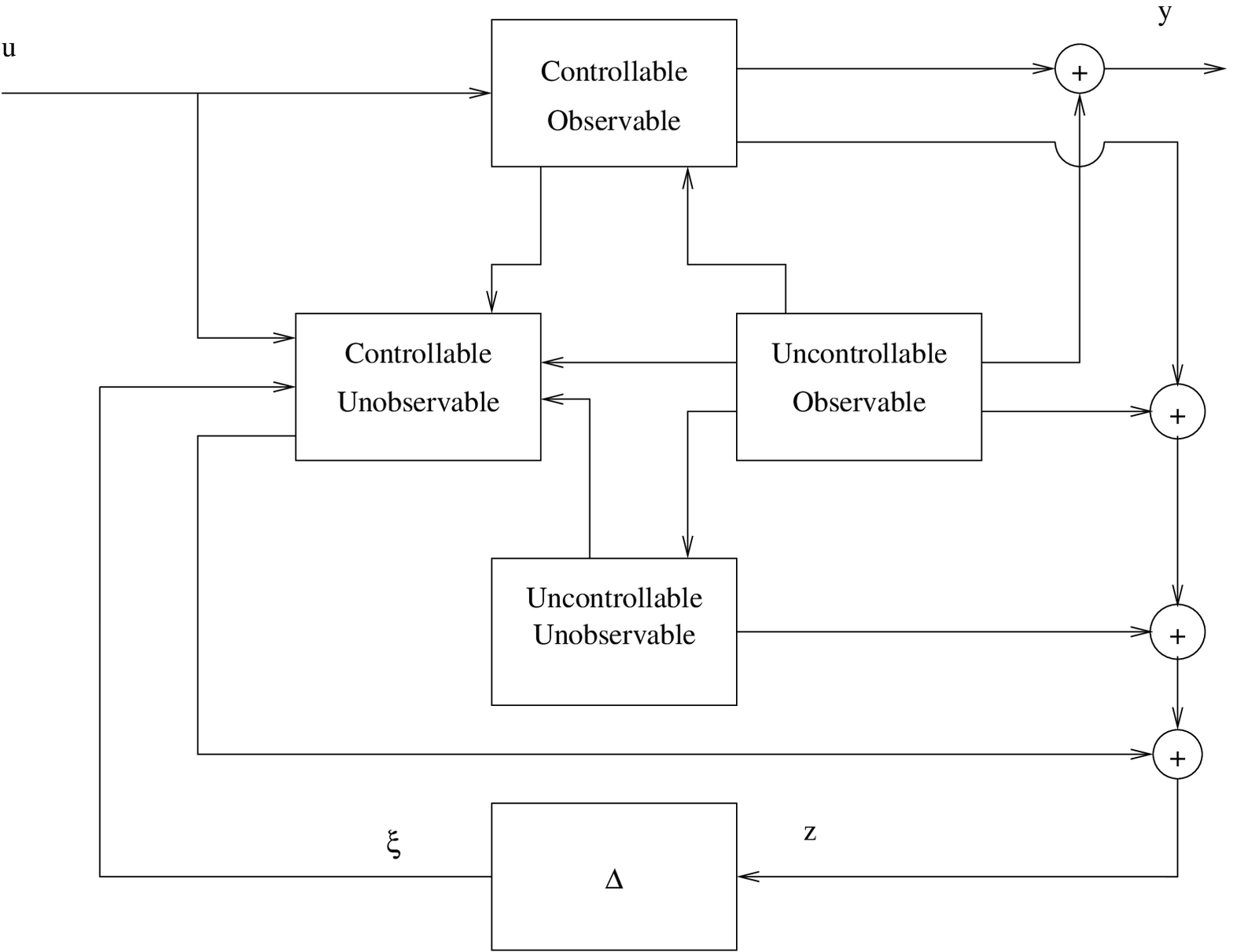}
\end{center}
\caption{Kalman decomposition for the uncertain
  system when $G(s)  \equiv 0$, $H(s) \not \equiv 0$.}
\label{F7}
\end{figure}

{\bf Case 3}
$G(s) \equiv 0$, $H(s) \not \equiv 0$. In this case, we  apply the
standard Kalman decomposition to the triple
$$(C_1,A,[B_1~B_2])$$
 to obtain
the situation as illustrated in the  block diagram shown in Fig.
 \ref{F7}. 

This situation corresponds to uncertainty only in the
controllable-unobservable block. Also there is uncertainty in the coupling between
controllable-observable block and each of the other blocks. 

\begin{figure}[htbp]
\begin{center}
\includegraphics[width=8cm]{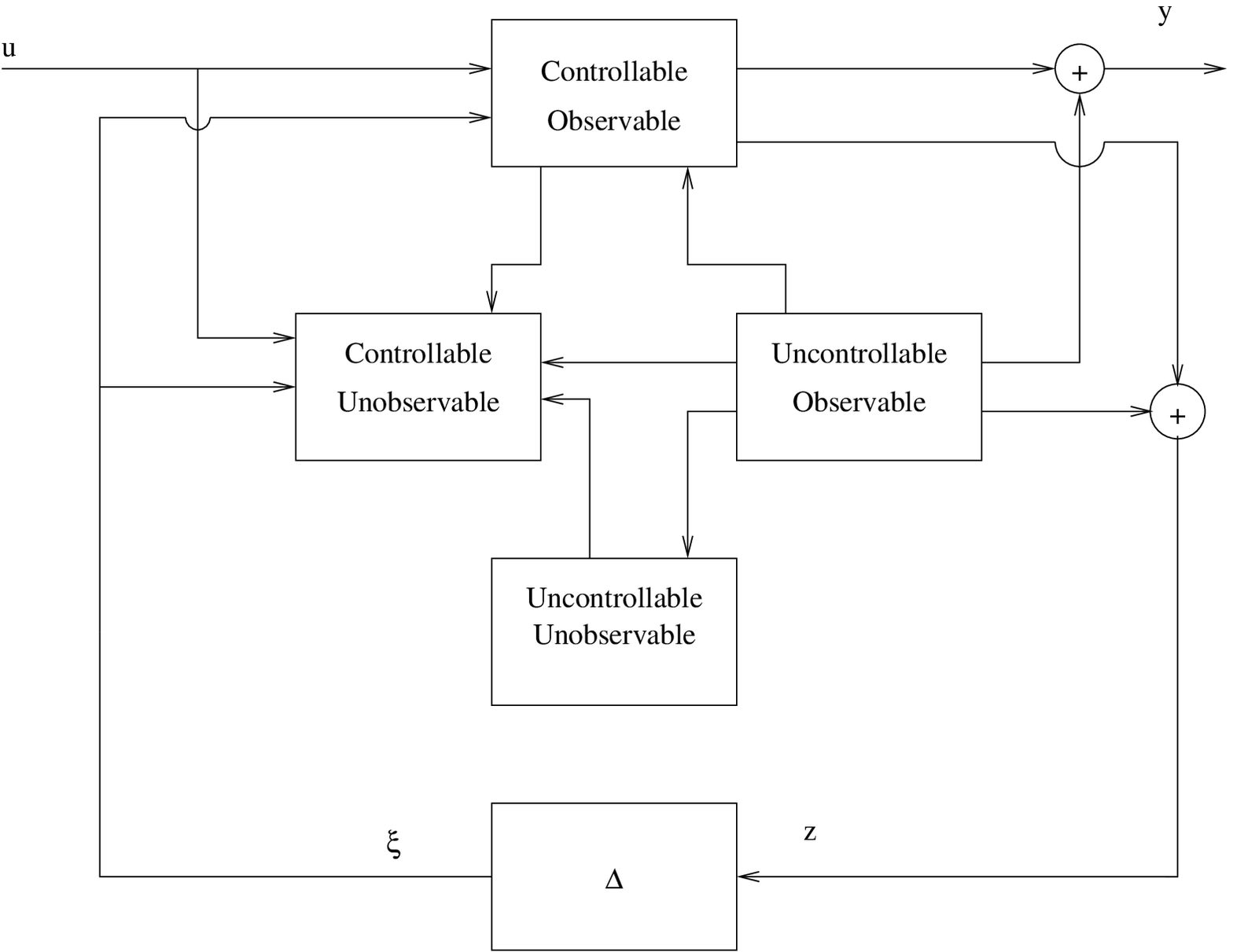}
\end{center}
\caption{Kalman decomposition for the uncertain
  system when $G(s)  \not \equiv 0$, $H(s) \not \equiv 0$.}
\label{F8}
\end{figure}

{\bf Case 4}
$G(s) \not \equiv 0$, $H(s) \not \equiv 0$. In this case, we  apply the
standard Kalman decomposition to the triple
$$(\left[\begin{array}{l}C_1\\C_2\end{array}\right],A,[B_1~B_2])$$
 to obtain
the situation as illustrated in the  block diagram shown in Fig.
 \ref{F8}. 

This situation corresponds to uncertainty only in the
controllable-observable block.
Also there is uncertainty in the coupling between
controllable-observable block and the uncontrollable-observable
block. Furthermore, there is uncertainty in the coupling between the controllable-observable
block and the controllable-unobservable block. As well, there is
uncertainty in the coupling between the uncontrollable-observable 
block and the controllable-unobservable block.

{\em Remark}
Note that each of the four cases considered above corresponds to
uncertainty only in one of the four blocks in the Kalman
decomposition. It might be conjectured that if structured uncertainty
was allowed then we could distribute the uncertainty blocks around the
four blocks in the Kalman decomposition rather than the current
requirement that the single uncertainty block corresponds to
uncertainty in one of the four blocks in the Kalman decomposition. 

\section{Illustrative Examples}
\label{example}
\subsection{Example 1}
\label{ex1}
In this example, we consider an uncertain system of the form (\ref{sys}), (\ref{int}) defined by the following matrices:
 \begin{eqnarray*}
 A &=& \left[\begin{array}{rr}
   -1.2838  &  0.3002\\
   -0.7603  & -0.2662
\end{array} \right]; ~
 B_1 = \left[\begin{array}{r}
    0.3911\\
    0.4348
 \end{array}\right];\nonumber \\
B_2 &=& \left[\begin{array}{r}
    0.7251    \\
    0.8062    
\end{array}\right];~
C_1 = \left[\begin{array}{rr}
    0.6534 &  -0.0908
 \end{array}\right];\nonumber \\
D_1&=&0; ~
 C_2 = \left[\begin{array}{rr}
-0.6190 &   0.5678
 \end{array}\right];~D_2=0.
 \end{eqnarray*}
This system is a modification of the system considered in the example of \cite{PET05B} to consider the case of unstructured uncertainty. We wish to determine if this uncertain system contains any states which are not possibly controllable in order to see if this uncertain system model can be replaced by an equivalent reduced dimension uncertain system model. We first calculate the transfer function 
$H(s) = C_1(sI-A)^{-1}B_1 + D_1 =  \frac{0.216 s + 0.1296}{s^2 + 1.55 s + 0.57} \not \equiv 0$. 
Hence, we apply Theorem \ref{T6} to this system and consider the uncontrollable states of the pair $(A,[B_1~ B_2])$; e.g., see \cite{AM06}. Indeed, the eigenvalues and corresponding left eigenvectors of the matrix $A$ are $\lambda_1 = -0.9500$, $\lambda_2 = -0.6000$, 
$x_1 =    \left[\begin{array}{r}
-0.9156\\
 0.4020
\end{array}\right]$, and
$x_2 =    \left[\begin{array}{r}
    0.7435\\
   -0.6687
\end{array}\right]$. Also, we have $B_1'x_2\approx B_2'x_2 \approx 0$. Hence (to the available numerical accuracy), $x_2$ is an uncontrollable state for the pair $(A,[B_1~ B_2])$. Hence using Theorem \ref{T6}, we can conclude that $x_2$ is not a possibly controllable state for this uncertain system.

\begin{figure}[htpb]
\begin{center}
\includegraphics[width=8cm]{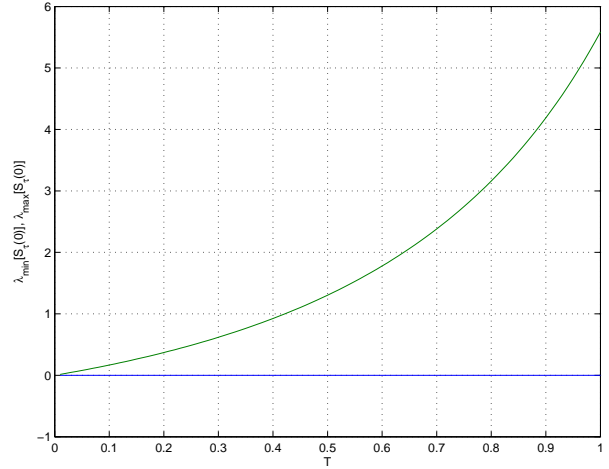}
\end{center}
\caption{$\lambda_{min}[S_\tau(0)]$ and $\lambda_{max}[S_\tau(0)]$
versus $T$ with $\tau =1$.}
\label{F9}
\end{figure}

We show that $x_2$ is not a possibly controllable state using Theorem \ref{T4}. Indeed, we let $\tau = 1$ and solve the Riccati differential equation (\ref{rdeS}) for different values of $T\in [0,1]$. A plot of the resulting eigenvalues of $S_\tau(0)$ versus $T$ is shown in Fig. \ref{F9}.
%
From this plot, we can see that the matrix $S_\tau(0)$ is singular for all $T\in [0,1]$. Furthermore, we find that $S_\tau(0)x_2 = 0$ for all  $T\in [0,1]$. Thus, using Theorem \ref{T4}, it follows that with $\tau = 1$,  $W_\tau(x_2,T) = \infty$ for all  $T\in [0,1]$. Hence, it follows from Definition \ref{D5} that the state $x_2$ is not (differentially) possibly
controllable. 

We now apply the Kalman decomposition to this uncertain system; e.g., see \cite{KAL63,KAL82,AM06}. Indeed,  if we apply the state space transformation $\tilde x = T x$ with $T = \left[ \begin{array}{rr}
   -0.7435 &   0.6687\\
    0.6687  &  0.7435
\end{array} \right]$ to this uncertain system, we obtain an uncertain system of the form (\ref{sys}), (\ref{int}) defined by:
 \begin{eqnarray*}
 \tilde A &=& \left[\begin{array}{rr}
   -0.6000 &  0.0000 \\
    1.0605 &  -0.9500
\end{array} \right]; ~
\tilde  B_1 = \left[\begin{array}{r}
   0.0000 \\
    0.5848
 \end{array}\right];\nonumber \\
\tilde B_2 &=& \left[\begin{array}{r}
    0.0000 \\
    1.0843     
\end{array}\right];~
\tilde C_1 = \left[\begin{array}{rr}
   -0.5465 &   0.3694
 \end{array}\right];\nonumber \\
\tilde D_1&=&0; ~
\tilde  C_2 = \left[\begin{array}{rr}
    0.8399 &   0.0082
 \end{array}\right];~\tilde D_2=0.
 \end{eqnarray*}
From this, the
control input $u$ and the uncertainty input $\xi$ do not affect the first state of this system. Thus, we can remove this state without changing the
input-output behavior of the system. This leads to a reduced dimension 
uncertain system  described by the state equations
\begin{eqnarray*}
\dot x &=&   -0.9500 x + 
    0.5848u +    1.0843 \xi;\nonumber \\
z &=&   0.3694 x;
y =    0.0082x
\end{eqnarray*}
and the averaged IQC (\ref{int}). 
\subsection{Example 2}
\begin{figure}[htpb]
\begin{center}
\psfrag{u}{$u$}
\psfrag{R1}{$R_1$}
\psfrag{R2}{$R_2$}
\psfrag{R3}{$R_3$}
\psfrag{C1}{$C_1$}
\psfrag{C2}{$C_2$}
\psfrag{V1}{$V_1$}
\psfrag{V2}{$V_2$}
\psfrag{y}{$y$}
\includegraphics[width=8cm]{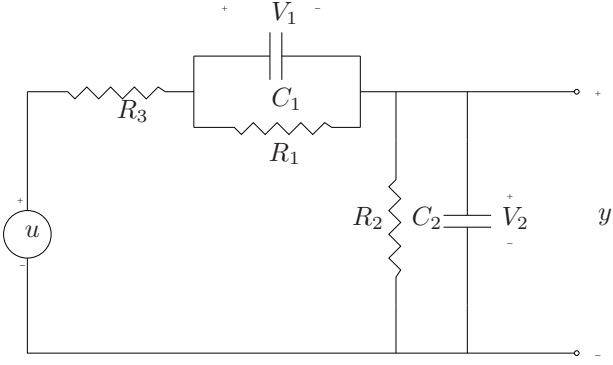}
\end{center}
\caption{Electrical circuit corresponding to Example 2.}
\label{F10}
\end{figure}

This example considers an uncertain system corresponding to the electrical circuit shown in Figure \ref{F10}. 
It is straightforward to derive the following state space model for this circuit:
\begin{eqnarray}
\label{ss_model1}
\left[\begin{array}{c}\frac{d V_1}{dt} \\ \frac{d V_2}{dt}\end{array}\right]
&=&\left[\begin{array}{cc} -\frac{1}{C_1}\left(\frac{1}{R_3}+\frac{1}{R_1}\right) & -\frac{1}{R_3 C_1}\\
-\frac{1}{R_3 C_2} & -\frac{1}{C_2}\left(\frac{1}{R_3}+\frac{1}{R_2}\right)\end{array}\right]
\left[\begin{array}{c}V_1 \\V_2 \end{array}\right]\nonumber \\
&&+ \left[\begin{array}{c}\frac{1}{C_1} \\\frac{1}{C_2} \end{array}\right]u;\nonumber \\
y&=& \left[\begin{array}{cc}0 & 1\end{array}\right]\left[\begin{array}{c}V_1 \\V_2 \end{array}\right].
\end{eqnarray}
In this example, we choose the  parameter values for the nominal to be $R_1 = 0.5 \Omega$, $R_2= 1.0 \Omega$, $R_3 = 0.5 \Omega$, $C_1 = 2.0 F$, and $C_2 = 1.0 F$. For these parameter values, the nominal system is not controllable. 

We now consider two cases of uncertain parameters for this system. In the first case, we find that all non-zero states of the system are possibly controllable and no reduced dimension model can be constructed using the Kalman decomposition of Section \ref{kalman}. In the second case, we find that there exist non-zero states of the system which are not possibly controllable. Then, we use the Kalman decomposition of Section \ref{kalman} to construct a model of order one which does not change the input-output behavior of the system. 

\noindent
{\bf Case 1.} In this case, we suppose that the conductance of the resistor $R_1$ is uncertain and we write
$
\frac{1}{R_1} = 2 + \Delta
$
where $|\Delta| \leq 1$. This leads to an uncertain system of the form (\ref{sys}) where 
\begin{eqnarray*}
A&=& \left[\begin{array}{cc}-2 & -1\\-2 & -3\end{array}\right];~~
B_1=\left[\begin{array}{c}0.5\\1\end{array}\right];~~
B_2=\left[\begin{array}{c}-0.5\\0\end{array}\right];\nonumber \\
C_1&=& \left[\begin{array}{cc}1 & 0 \end{array}\right];~~
D_1= 0;~~
C_2= \left[\begin{array}{cc}0 & 1 \end{array}\right];~~
D_2=0;
\end{eqnarray*}
and $\xi = \Delta z$. Since $|\Delta| \leq 1$, it follows that the averaged IQC (\ref{int}) will be satisfied. For this uncertain system, we calculate the transfer functions $G(s)$ and $H(s)$ as
\begin{eqnarray*}
G(s) &=&  \frac{1}{s^2 + 5 s + 4}\not \equiv 0;~~
H(s) = \frac{0.5 s + 0.5}{s^2 + 5 s + 4} \not \equiv 0.
\end{eqnarray*}
For this uncertain system, the pair $(A,B_1)$ is not controllable. However, the pair $(A,[B_1~ B_2])$ is controllable. Thus, it follows from Theorem \ref{T6} that the system has no states which are not differentially possibly controllable. Also, the pair 
$(\left[\begin{array}{l}C_1\\C_2\end{array}\right],A)$ is observable and hence, using Case 4 of the Kalman decompositions considered in Section \ref{kalman}, we cannot construct an equivalent reduced dimension uncertain system corresponding to this uncertain system. 

Note that the example considered in this case is such that the nominal system is not controllable, but the uncertain system becomes controllable for non-zero values of the uncertain parameter $\Delta$. If we change the parameter $C_1$ to $C_1 = 1$, we obtain an uncertain system for which the nominal system is controllable but for which the system becomes uncontrollable for one value of the uncertain parameter ($\Delta = -1$). 

\noindent
{\bf Case 2.} In this case, we suppose that the conductance of the resistor $R_3$ is uncertain and we write
$
\frac{1}{R_3} = 2 + \Delta
$
where $|\Delta| \leq 1$. This leads to an uncertain system of the form (\ref{sys}) where 
\begin{eqnarray*}
A&=& \left[\begin{array}{cc}-2 & -1\\-2 & -3\end{array}\right];~~
B_1=\left[\begin{array}{c}0.5\\1\end{array}\right];~~
B_2=\left[\begin{array}{c}-0.5\\-1\end{array}\right];\nonumber \\
C_1&=& \left[\begin{array}{cc}1 & 1 \end{array}\right];~~
D_1= 0;~~
C_2= \left[\begin{array}{cc}0 & 1 \end{array}\right];~~
D_2=0;
\end{eqnarray*}
and $\xi = \Delta z$. For this uncertain system, we calculate the transfer functions $G(s)$ and $H(s)$ as
\begin{eqnarray*}
G(s) &=&  \frac{-s -1 }{s^2 + 5 s + 4}\not \equiv 0;~~
H(s) = \frac{1.5 s + 1.5}{s^2 + 5 s + 4} \not \equiv 0.
\end{eqnarray*}
For this uncertain system, the pair $(A,[B_1~ B_2])$ is not controllable. Thus, it follows from Theorem \ref{T6} that the system has non-zero states which are not differentially possibly controllable. Also, the pair 
$(\left[\begin{array}{l}C_1\\C_2\end{array}\right],A)$ is observable. We now construct the Kalman decomposition for this system as in  Case 4 of  Section \ref{kalman}. Indeed, we apply a state space transformation $\tilde x = T x$ with $T = \left[ \begin{array}{rr}
   -0.8944    & 0.4472\\
   -0.4472   & -0.8944
\end{array} \right]$ to this uncertain system to  obtain an uncertain system of the form (\ref{sys}), (\ref{int}) defined by:
 \begin{eqnarray*}
 \tilde A &=& \left[\begin{array}{rr}
   -1.0000   & 0.0000\\
   -1.0000   &-4.0000
\end{array} \right]; ~
\tilde  B_1 = \left[\begin{array}{r}
    0.0000 \\
   -1.1180
 \end{array}\right];\nonumber \\
\tilde B_2 &=& \left[\begin{array}{r}
   -0.0000\\
    1.1180
\end{array}\right];~
\tilde C_1 = \left[\begin{array}{rr}
   -0.4472  & -1.3416
 \end{array}\right];\nonumber \\
\tilde D_1&=&0; ~
\tilde  C_2 = \left[\begin{array}{rr}
    0.4472 &  -0.8944
 \end{array}\right];~\tilde D_2=0.
 \end{eqnarray*}
From this, the 
control input $u$ and the uncertainty input $\xi$ do not affect the first state of this system. Thus, we can remove this state without changing the
input-output behavior of the system. This leads to a reduced dimension 
uncertain system  described by the state equations
\begin{eqnarray*}
\dot x &=&   -4.0x -
    1.1180u +    1.1180 \xi;\nonumber \\
z &=& -1.3416  x;
y =    -0.8944 x
\end{eqnarray*}
and the averaged IQC (\ref{int}). 

\section{Conclusions and Future Research}
\label{conclusion}
The results of this paper have led to a geometric characterization of
the notion of possible controllability for a class of uncertain linear
systems. These results
combined with a corresponding geometric 
characterization of the notion of robust unobservability have allowed
us to present a complete Kalman decomposition for uncertain systems. 

Possible areas of future research motivated by the results of this
paper include 
extending the results of the paper to the case of structured uncertainty
subject to multiple IQCs.


\end{document}